\documentclass[twocolumn,showpacs,preprintnumbers,nofootinbib,amsmath,amssymb]{revtex4}

\usepackage{euscript,amssymb,amsmath}

\usepackage{graphicx}
\usepackage{epsfig}

\newcommand{\be}[1]{\begin{equation}\label{#1}}
\newcommand{\ee}{\end{equation}}
\newcommand{\ba}[1]{\begin{eqnarray}\label{#1}}
\newcommand{\ea}{\end{eqnarray}}
\newcommand{\rf}[1]{(\ref{#1})}
\newcommand{\nn}{\nonumber}

\begin{document}

\title{Observational constraints on Kaluza-Klein models \\ with $d$-dimensional spherical compactification}

\author{Maxim Eingorn}\email{maxim.eingorn@gmail.com}   \author{Seyed Hossein Fakhr}\email{hxf1_2011@yahoo.com}   \author{Alexander Zhuk}\email{ai.zhuk2@gmail.com}

\affiliation{Astronomical Observatory, Odessa National University, Street Dvoryanskaya 2, Odessa 65082, Ukraine\\}

\begin{abstract}
We investigate Kaluza-Klein models in the case of spherical compactification of the internal space with an arbitrary number of dimensions. The gravitating source has the
dust-like equation of state in the external/our space and an arbitrary equation of state (with the parameter $\Omega$) in the internal space. We get the perturbed (up to
$O\left(1/c^2\right)$) metric coefficients. For the external space, these coefficients consist of two parts: the standard general relativity expressions plus the
admixture of the Yukawa interaction. This admixture takes place only for some certain condition which is equivalent to the condition for the internal space
stabilization. We demonstrate that the mass of the Yukawa interaction is defined by the mass of the gravexciton/radion. In the Solar system, the Yukawa mass is big
enough for dropping the admixture of this interaction and getting good agreement with the gravitational tests for any value of $\Omega$. However,  the gravitating body
acquires the effective relativistic pressure in the external space which vanishes only in the case of tension $\Omega=-1/2$ in the internal space.
\end{abstract}


\maketitle

\vspace{.5cm}


\section{\label{sec:1}Introduction}

The idea of multidimensionality of spacetime is one of the most intriguing hypotheses of past and present centuries. For example, one of the main tasks of the LHC is to
search for evidence of extra dimensions. However, there is a large variety of multidimensional models. It is obvious that the search for extra dimensions is most
effective if we know which of these models are viable. In other words, we need to know which of these models do not contradict the observational data. The known
gravitational experiments (the perihelion shift, the deflection of light, the time delay of radar echoes) in the Solar system are good filters to screen out non-physical
theories. It is well known that the weak field approximation is enough to calculate the corresponding formulae for these experiments \cite{Landau}. For example, in the
case of general relativity these formulae demonstrate excellent agreement with the experimental data. In our previous papers  we investigated the popular Kaluza-Klein
models with toroidal compactification of the internal spaces.  As we have shown, to be at the same level of agreement with the gravitational tests as general relativity,
the gravitating masses should have tension in the internal spaces. This is true for both linear \cite{EZ4,EZ5} and nonlinear $f(R)$ \cite{EZnonlin,EZf(R)solitons}
models. Unfortunately, the physically reasonable models with the dust-like equation of state $p=0$ in all spaces contradict these tests \cite{EZ3}. It happens because of
the fifth force generated by variations of the internal space volume \cite{EZ6}. For the proper value of tension, the internal space volume is fixed and the fifth force
is absent. For example, it takes place for the latent solitons and black strings/branes as their particular cases.

However, up to now we are not aware of the physical meaning of tension for the ordinary astrophysical objects similar to our Sun\footnote{For black strings and black
branes, the notion of tension is defined, e.g., in \cite{TF} and it follows from the first law for black hole spacetimes \cite{TZ,HO,TK}. Obviously, our Sun is not a
relativistic astrophysical object. Therefore, to calculate motion of a test body in the vicinity of the Sun, there is no need to take into account the relativistic
properties of black holes (e.g., the presence of the horizon). It is sufficient to use  the corresponding black strings/branes metrics in the weak field limit.}.
Therefore, we continued to search for models that satisfy the gravitational tests, being free from this physically unclear property. For this purpose, we have considered
Kaluza-Klein models with spherical compactification of the internal space, being a two-sphere \cite{ChEZ1,ChEZ2}. Here, we have shown that
the conformal variation of the volume of the two-sphere generates the Yukawa-type admixture to the metric coefficients. The characteristic range of the Yukawa
interaction for this model is proportional to the scale factor of the two-sphere (the radius of the two-sphere): $\lambda \sim a$. On the one hand, the sizes of the
extra dimensions in the Kaluza-Klein models are bounded by the recent collider experiments. On the other hand, there are strong restrictions on the Yukawa parameter
$\lambda$ from the inverse square law experiments. According to both of these limitations, $\lambda$ is by many orders of magnitude smaller than  the radius of the Sun.
Then, with very high accuracy, we can drop the admixture of the Yukawa terms to the metric coefficients at distances greater than the radius of the Sun, and we achieve
good agreement with the gravitational tests for the considered model. In this model, the gravitating mass has the dust-like equation of state in all spatial dimensions,
i.e. tension is absent.

The following question arises: how general is this result? Does it depend on the number of the extra dimensions? The point is that the critical number of spacetime
dimensions is 10 for superstring theories and 11 for M-theory. Therefore, in the present paper, the internal space is an arbitrary $d$-dimensional sphere. We also do not
limit ourselves with the dust-like equation of state of the gravitating source in all spaces. We assume that the gravitating mass has this equation of state $\hat p_0=0$
in the external/our space, and the equation of state $\hat p_1\approx\Omega\hat\rho c^2$ in the internal space, where the parameter $\Omega$ is arbitrary. It turns out
that this generalization plays an important role. We show that the considered models with the $d$-dimensional internal sphere can satisfy the gravitational tests for any
$d\ge 2$ and for an arbitrary value $\Omega$ of the parameter of the equation of state in the internal space, including the dust-like value $\Omega=0$. We also get the
internal space stability condition. For models which satisfy this condition, the internal space conformal excitations result in the Yukawa-type correction terms to the
metric coefficients. Moreover, we demonstrate that the Yukawa mass is defined by the mass of the gravexciton/radion \cite{exci}. Therefore, for sufficiently large Yukawa
mass (that takes place in the Solar system) the considered models satisfy the gravitational tests for an arbitrary value of $\Omega$. However, we show that the
gravitating body acquires the effective relativistic pressure in the external space. Obviously, it is not the case for ordinary astrophysical objects similar to our Sun.
Such pressure disappears only in the case of the bare tension $\Omega =-1/2$ in the internal space. Therefore, to be in agreement with observations, we again should
include tension in the internal space as it takes place for the toroidal compactification.

The paper is organized as follows. In Sec. II we define the background solution in the case of spherical compactification of the internal space with an arbitrary number
of dimensions. Then, we perturb this solution by a gravitating mass with an arbitrary equation of state in the internal space, and obtain the metric correction terms
from the Einstein equations. Here, we show that the conformal variations of the internal space result in the Yukawa interaction with the mass defined by the mass of
radion. In Sec. III we show that the gravitating body acquires the effective relativistic pressure in the external space which vanishes only in the case of tension in
the internal space. The main results are summarized in the concluding Sec. IV.

\section{\label{sec:2}Background solution and perturbations}

Before we consider the gravitational field produced by the gravitating mass, we need to create an appropriate background metrics. Such metrics is defined on the product
manifold $M = M_4\times M_d$, where $M_4$ describes external four-dimensional flat spacetime and $M_d$ corresponds to the $d$-dimensional internal space which is a
sphere with the radius (the internal space scale factor) $a$, and should have the form
\be{2.1}
 ds^2=c^2 dt^2-dx^2-dy^2-dz^2 + \sum_{\mu=4}^D g_{\mu\mu}d\xi_{\mu}^2\, ,
\ee
where
\ba{2.2}
g_{DD}&=&-a^2\, ,\nn\\
g_{\mu\mu}&=&-a^2\prod_{\nu=\mu+1}^{D}\sin^2\xi_{\nu}\, ,\; \; \mu=4,\ldots ,D-1\, ,\ea
and $D=3+d$ is the total number of spatial dimensions. To create such metrics with the curved internal space, we have to introduce background matter with the
energy-momentum tensor
\be{2.3}
\bar T_{ik}=\left\{
\begin{array}{cc}
\left( \frac{d(d-1)}{2\kappa a^2}-\Lambda_\mathcal{D} \right) g_{ik} & \mbox{for   } \, i,k=0,...,3;\\
\\
\left(\frac{(d-1)(d-2)}{2\kappa a^2} - \Lambda_\mathcal{D} \right)g_{ik} & \mbox{for   } \, i, k=4,5,\ldots ,D.
\end{array} \right . \quad
\ee
These components of the energy-momentum tensor can be easily got from the Einstein equation
\be{2.4} \kappa \bar T_{ik} = R_{ik}-\frac{1}{2} R g_{ik}-\kappa\Lambda_{\mathcal{D}} g_{ik}\,  \ee
for the background metrics \rf{2.1}.
 Here, $\kappa \equiv 2S_D\tilde G_{\mathcal{D}}/c^4$, $S_D=2\pi^{D/2}/\Gamma (D/2)$ is the total solid angle (the surface area of the $(D-1)$-dimensional sphere of a
unit radius) and $\tilde G_{\mathcal{D}}$ is the gravitational constant in $(\mathcal{D}=D+1)$-dimensional spacetime. We also include in the model a bare
multidimensional cosmological constant $\Lambda_{\mathcal{D}}$. To get these components of the energy-momentum tensor, we took into account that the only non-zero
Ricci-tensor components are $R_{\mu\mu}=-[(d-1)/a^2]g_{\mu\mu},\; \mu=4,\ldots ,D$, and the scalar curvature is $R=-d(d-1)/a^2$.

It can be easily seen that the expression \rf{2.3} can be written in the form of the energy-momentum tensor of a perfect fluid:
\be{2.5} \bar T^i_k = \mbox{diag}\left(\bar{\varepsilon}, -\bar{p}_0,-\bar{p}_0,-\bar{p}_0,\underbrace{-\bar{p}_1,\ldots ,-\bar{p}_1}_{d \ times}\right)\, , \ee
where the energy density and pressures of the background matter in the external and internal spaces are respectively
\ba{2.6}
\bar{\varepsilon} &\equiv& \frac{d(d-1)}{2\kappa a^2}-\Lambda_{\mathcal{D}}\, ,\nn\\
\bar{p}_0 &\equiv& -\frac{d(d-1)}{2\kappa a^2}+\Lambda_{\mathcal{D}}\, ,\\
\bar{p}_1 &\equiv& -\frac{(d-1)(d-2)}{2\kappa a^2}+\Lambda_{\mathcal{D}}\, \nn.
\ea
Therefore, we get the vacuum-like equation of state in the external space:
\be{2.7}
\bar{p}_0=\omega_0 \bar \varepsilon\, , \quad \omega_0 =-1,
\ee
but the equation of state in the internal space is not fixed:
\ba{2.8}
&{}&\bar{p}_1=\omega_1\bar \varepsilon \, , \nn\\
&{}& \omega_1 = \frac{2\Lambda_{\mathcal{D}}\kappa a^2-(d-1)(d-2)}{d(d-1)-2\Lambda_{\mathcal{D}}\kappa a^2}\, \, ,
\ea
i.e. $\omega_1$ is arbitrary.  Choosing different values of $\omega_1$ (with fixed $\omega_0 = -1$), we can simulate different forms of matter. For example, $\omega_1=1$
corresponds to the monopole form-fields (the Freund-Rubin scheme of compactification \cite{FR})\footnote{In our case, they are $d$-forms (see, e.g., Eqs. (2.9) and (5.1)
in \cite{Zhuk}).}, and for the Casimir effect we have $\omega_1=4/d$ \cite{exci}. It is worth noting that the parameter $\omega_1$ can be positive only in the presence
of a positive bare cosmological constant $\Lambda_{\mathcal{D}}$. Moreover, it takes place only if $d-2<2\Lambda_{\mathcal{D}}\kappa a^2/(d-1)<d$. In contrast to the
model with the two-dimensional sphere in \cite{ChEZ2}, the parameter $\omega_1$ does not disappear in the case of vanishing $\Lambda_{\mathcal{D}}$ for $d\geq 3$.

Now, we perturb our background ansatz by a static point-like massive source with non-relativistic rest mass density $\hat \rho$. We suppose that the matter source is
uniformly smeared over the internal space \cite{EZ2}. Hence, multidimensional $\hat \rho$ and three-dimensional $\hat \rho_3$ rest mass densities are connected as
follows: $\hat \rho=\hat \rho_3({\bf r}_3)/V_{int}$ where $V_{int}=\left[2\pi^{(d+1)/2}/\Gamma((d+1)/2)\right]a^d$ is the surface area of  the $d$-dimensional sphere of
the radius $a$ (see, e.g., \cite{EZ3,EZ2}). In the case of a point-like mass $m$, $\hat \rho_3(r_3)=m\delta ({\bf r}_3)$, where $r_3=|{\bf r}_3|=\sqrt{x^2+y^2+z^2}$. In
the non-relativistic approximation the energy density
of the point-like mass is $\hat{T}^0_{0}\approx \hat \rho c^2$ and up to linear in perturbations terms $\hat{T}_{00}\approx \rho c^2$. Inasmuch as the gravitating mass
is at rest in the external space, it has the dust-like equation of state $\hat p_0=0\; \Rightarrow \; \hat{T}^1_{1}=\hat{T}^2_{2}=\hat{T}^3_{3}=0$ in our dimensions.
However, it may have the non-zero equation of state  $\hat p_1 \approx \Omega \hat \rho c^2\; \Rightarrow \; \hat T_{\mu}^{\mu}\approx -\Omega \hat \rho c^2 \, ,\; \mu
=4,5,\ldots ,D$ in the internal space. We shall see that this generalization has important consequences. All other components of the energy-momentum tensor of the
gravitating mass are equal to zero.

Concerning the energy-momentum tensor of the background matter, we suppose that perturbation does not change the equations of state in the external and internal spaces,
i.e. $\omega_0$ and $\omega_1$ are constants. For example, if we had monopole form-fields ($\omega_0=-1,\,\omega_1=1$) before the perturbation, the same type of matter
we shall have after the perturbation. Therefore, the energy-momentum tensor of the perturbed background is
\be{2.9}
\tilde T_{ik}\approx\left\{
   \begin{array}{cc}
   \left(\bar\varepsilon +\varepsilon^1\right) g_{ik}, & i,k=0,...,3;\\
   \\
   -\omega_1 \left(\bar\varepsilon +\varepsilon^1\right) g_{ik}, & i, k=4,5,\ldots ,D\, ,
   \end{array}
\right.
\ee
where the correction $\varepsilon^1$ is of the same order of magnitude as the perturbation $\hat \rho c^2$.

We suppose that the perturbed metrics preserves its diagonal form. Obviously, the off-diagonal coefficients $g_{0\alpha}, \, \alpha =1,\ldots ,D$, are absent for the
static metrics. It is also clear that in the case of the uniformly smeared (over the internal space) gravitating mass, the perturbed metric coefficients (see functions
$A,B,C,D$ and $G$ below) depend only on $x,y,z$ \cite{EZ2}, and the metric structure of the internal space does not change, i.e.
$g_{\mu\mu}=g_{DD}\prod_{\nu=\mu+1}^{D}\sin^2\xi_{\nu}\, , \mu=4,\ldots ,D-1$. The latter statement can be proved, e.g., in the weak field approximation from the
Einstein equations (see appendix B in \cite{ChEZ2}). It is also easy to show that in this case the spatial part of the external metrics can be diagonalized by coordinate
transformations.
Therefore, the perturbed metrics reads
\ba{2.10} ds^2&=&Ac^2dt^2+Bdx^2+Cdy^2+Ddz^2\nn \\
&+&G\left[\sum_{\mu=4}^{D-1}\left( \prod_{\nu=\mu+1}^{D}\sin^2\xi_{\nu} \right)d\xi_{\mu}^2+d\xi_D^2\right]
\ea
with
\ba{2.11}
&{}& A \approx 1+A^{1}({r}_3),\quad B \approx -1+B^{1}({r}_3)\, ,\nn\\
&{}& C \approx -1+C^{1}({r}_3),\quad D \approx -1+D^{1}({r}_3)\, ,\nn\\
&{}& G \approx -a^2+G^{1}({r}_3)\, .\ea
All metric perturbations $A^1,B^1,C^1,D^1,G^1$ are of the order of $\varepsilon^1$. To find these corrections as well as the background matter perturbation
$\varepsilon^1$, we should solve the Einstein equation
\be{2.12}
R_{ik}=\kappa\left( T_{ik}-\frac{1}{2+d}T g_{ik} -\frac{2}{2+d} \Lambda_{\mathcal{D}} g_{ik}\right)\, ,
\ee
where the energy-momentum tensor $T_{ik}$ is the sum of the perturbed background $\tilde T_{ik}$ \rf{2.9} and the energy-momentum tensor of the perturbation $\hat
T_{ik}$. First, we would like to note that the diagonal components of the Ricci tensor for the metrics \rf{2.10} up to linear
terms $A^1,B^1,C^1,D^1,G^1$ are
\ba{2.13}
R_{00} &\approx& \frac{1}{2}\triangle_3 A^1\, ,\nn \\
R_{11}&\approx&\frac{1}{2}\triangle_3 B^1
+\frac{1}{2}\left(-A^1-B^1+C^1+D^1+d\frac{G^1}{a^2}\right)_{xx}\, ,\nn \\
R_{22}&\approx&\frac{1}{2}\triangle_3 C^1
+\frac{1}{2}\left(-A^1+B^1-C^1+D^1+d\frac{G^1}{a^2}\right)_{yy}\, ,\nn \\
R_{33}&\approx&\frac{1}{2}\triangle_3 D^1
+\frac{1}{2}\left(-A^1+B^1+C^1-D^1+d\frac{G^1}{a^2}\right)_{zz}\, ,\nn \\
R_{DD}&\approx& d-1+\frac{1}{2}\triangle_3 G^1\, , \ea
where $\triangle_3=\partial^2/\partial x^2+\partial^2/\partial y^2+\partial^2/\partial z^2$ is the three-dimensional Laplace operator. Additionally, for the static
metrics \rf{2.10}, where the coefficients $A,B,C,D$ and $G$ depend only on $x,y,z$, there is the following relation:
\be{2.14} R_{\mu\mu} = R_{DD}\prod_{\nu=\mu+1}^{D}\sin^2\xi_{\nu}\, ,\quad \mu=4,\ldots ,D-1\, . \ee
Concerning the off-diagonal components of the Ricci tensor, they should be equal to zero according to the Einstein equation \rf{2.12}. Taking into account the relations
$g_{\mu\mu}=g_{DD}\prod_{\nu=\mu+1}^{D}\sin^2\xi_{\nu}\, , \mu=4,\ldots ,D-1$, and that the metric coefficients $A,B,C,D$ and $G$ depend only on $x,y,z$, we can easily
verify that all off-diagonal components are identically equal to zero except the components $R_{12}, R_{13}$ and $R_{23}$. Equating these components to zero, we obtain
the following relations between the metric coefficients:
\be{2.15}
B^1=C^1=D^1
\ee
and
\be{2.16}
A^1-B^1-\frac{d}{a^2}G^1=0\, ,
\ee
which demonstrate that the expressions in brackets for 11, 22 and 33 components in \rf{2.13} vanish. Therefore, in the weak field limit the Einstein equation \rf{2.12}
is reduced to the following system of equations:
\ba{2.17}
\triangle_3 A^1 &=&\frac{1+d+d\Omega}{1+d/2}\kappa\hat\rho c^{2}+\frac{d-2+d\omega_{1}}{1+d/2}\kappa\varepsilon^{1}\, ,\\
\triangle_3 B^1 &=&\frac{1-d\Omega}{1+d/2}\kappa\hat\rho c^{2}-\frac{d-2+d\omega_{1}}{1+d/2}\kappa\varepsilon^{1}\, ,\label{2.18}\\
\triangle_3 G^1 &=&\frac{1+2\Omega}{1+d/2}\kappa a^{2}\hat\rho c^{2}+\frac{4+2\omega_{1}}{1+d/2}\kappa a^{2}\varepsilon^{1}\nn \\
&-&\frac{2(d-1)}{a^2} G^1\label{2.19}\, .
\ea
From these equations and the condition \rf{2.16} we obtain the connection between $G^1$ and $\varepsilon^1$:
\be{2.20}
\kappa\varepsilon^{1}=\frac{d(d-1)}{2  a^{4}}G^1\, .
\ee
Then, the system of Eqs. \rf{2.17}-\rf{2.19} reads:
\ba{2.21}
&\triangle_3& \left(A^1-\frac{d}{2a^2}G^1\right)=\kappa \hat\rho c^{2}=\frac{8\pi G_N}{c^2}\hat\rho_3\, ,\\
&\triangle_3& \left(B^1+\frac{d}{2a^{2}}G^1\right)=\kappa \hat\rho c^{2}=\frac{8\pi G_N}{c^2}\hat\rho_3\label{2.22}\, ,\\
&\triangle_3& G^1 -\lambda^{-2}G^1=\frac{2(1+2\Omega)}{2+d}\kappa a^{2}\hat\rho c^{2}\nn\\
&=&a^2\frac{2(1+2\Omega)}{2+d}\frac{8\pi G_N}{c^2}\hat\rho_3\, ,\label{2.23}
\ea
where
\be{2.24} \lambda^2\equiv \frac{(d+2)a^2}{2(d-1)(d-2+d\omega_1)} \ee
and we introduced the Newton gravitational constant
\be{2.25}
4\pi G_N = \frac{S_D \tilde G_\mathcal{D}}{V_{int}}\, .
\ee
Let us consider now the point-like (in the external space) approximation for the gravitating objects: $\hat \rho_3(r_3)=m\delta ({\bf r}_3)$. The generalization of the
obtained results to the case of extended compact objects is obvious. It is well known that to get the physically reasonable solution of \rf{2.23} with the boundary
condition $G^1 \to 0$ for $r_3 \to +\infty$ the parameter $\lambda^2$ should be positive, i.e. the equation of state parameter $\omega_1$ should satisfy the condition
\be{2.26}
\omega_1> -1+\frac{2}{d}\, ,
\ee
which allows also negative values of $\omega_1$. From Eq. \rf{2.8} we can get the corresponding restrictions for the bare cosmological constant:
\be{2.27} 0<\frac{2\Lambda_{\mathcal{D}}\kappa a^2}{d-1}<d\, . \ee
This inequality relaxes the condition of the positiveness of $\omega_1$. Then, the Eqs. \rf{2.21}-\rf{2.23} have solutions
\ba{2.28}
A^1&=&\frac{2\varphi_N}{c^2}+\frac{d}{2a^{2}}G^1\, ,\\
B^1&=&\frac{2\varphi_N}{c^2}-\frac{d}{2a^{2}}G^1\, ,\label{2.29}\\
G^1&=&a^2\frac{4\varphi_N}{(2+d)c^2}(1+2\Omega)\exp\left(-\frac{r_3}{\lambda}\right)\, ,\label{2.30}
\ea
where the Newtonian potential $\varphi_N=-G_N m/r_3$. It is well known that the metric correction term $A^1\sim O\left(1/c^2\right)$ describes the non-relativistic
gravitational potential: $A^1=2\varphi/c^2$. Therefore, this potential acquires the Yukawa correction term:
\be{2.31} \varphi = \varphi_N\left[1+\frac{d}{2+d}(1+2\Omega)\exp\left(-\frac{r_3}{\lambda}\right)\right]\, . \ee

The inequalities \rf{2.26} and \rf{2.27} provide the condition of the internal space stabilization. Obviously, the monopole form-field ansatz with $\omega_1=1$ satisfies
this condition. Let us consider this example in more detail. From the fine-tuning relation \rf{2.8} we obtain $2\Lambda_{\mathcal{D}}\kappa a^2=(d-1)^2$. Precisely this
quantity provides a zero value of the effective four-dimensional cosmological constant (see Eq. (5.7) in \cite{Zhuk} where we should make the substitutions $\Lambda_{D}
\to \Lambda_{\mathcal{D}}\kappa ,\, d_1 \to d ,\, D \to 4+d $ and where $\tilde R_1=d(d-1)/a^2,\, d_0=3, D_0=4$). It is clear that in our case with flat background
external spacetime, the effective four-dimensional cosmological constant should vanish. Additionally, this value of $\Lambda_{\mathcal{D}}$ satisfies the stability
condition (5.15) in \cite{Zhuk}. Moreover, the gravexciton/radion mass squared (5.12) (with substitution (5.11)) exactly coincides with the Yukawa mass squared
$m_{Yu}^2=\lambda^{-2}=4(d-1)^2/[(d+2)a^2]\equiv m_{exci}^2$. Therefore, we arrived at the very natural conclusion that the Yukawa mass is defined by the mass of the
gravexcitons/radions.

For reasonable values of the equation of state parameter $|\Omega|\sim O(1)$, the Yukawa parameter $\alpha$ (the prefactor in front of the exponential function) is also
of the order of 1. Then, the inverse square law experiments restrict the characteristic range of the Yukawa interaction: $\lambda \lesssim 10^{-3}$cm. Obviously, for the
above-mentioned gravitational experiments in the Solar system $r_3\gtrsim r_{\odot}\sim 7\times 10^{10}$cm, and the ratio $r_3/\lambda \gtrsim 10^{14}$. On the other
hand, the collider experiments also restrict the sizes of the extra dimensions for Kaluza-Klein models: $a\lesssim 10^{-17}$cm. Since $\lambda \sim a$, then $r_3/\lambda
\gtrsim 10^{28}$. It is clear that for such ratios we can drop the Yukawa correction terms in Eqs. \rf{2.28} and \rf{2.29}. Therefore, the post-Newtonian parameter
$\gamma=B^1/A^1$ is equal to 1 with extremely high accuracy for any value of $\Omega$ including the most physically reasonable case of the dust-like value $\Omega=0$.
The case $\Omega=-1/2$ is a special one, and we consider it below. Thus we arrive at the concordance with the gravitational tests (the deflection of light and the time
delay of radar echoes) for our model.


\section{\label{sec:3}Effective energy-momentum tensor for gravitating mass}

So, at first glance, it seems that we have found a model, which, on the one hand, satisfies the gravitational experiments and, on the other hand, may not contain
tension in the internal space. However, let us examine in detail the energy-momentum tensor of the gravitating mass. As follows from Eqs. \rf{2.20} and \rf{2.30}, the
background perturbation $\varepsilon^1$ is localized around the gravitating object and falls exponentially with the distance $r_3$ from this mass. Therefore, the bare
gravitating mass is covered by this "coat".  For an external observer, this coated gravitating mass is characterized by the effective energy-momentum tensor with the
following nonzero components:
\ba{3.1}
&{}&T_{0}^{0(eff)}\approx \varepsilon^1+\hat\rho({\bf r}_3)c^2=\\
&{}&-(1+2\Omega)\frac{d(d-1)mc^2}{4(2+d)\pi V_{int}a^2}\frac{1}{r_3}\exp\left(-\frac{r_3}{\lambda}\right)+\frac{mc^2\delta({\bf
r}_3)}{V_{int}}\, ,\nn\\
&{}&T_{\alpha}^{\alpha (eff)}\approx \varepsilon^1\label{3.2}=\\
&{}&-(1+2\Omega)\frac{d(d-1)mc^2}{4(2+d)\pi V_{int}a^2}\frac{1}{r_3}\exp\left(-\frac{r_3}{\lambda}\right),\quad\alpha=1,2,3\, ,\nn\\
&{}&T_{\mu}^{\mu(eff)}\approx-\omega_1\varepsilon^1-\Omega\hat\rho({\bf r}_3) c^2=\label{3.3}\nn\\
&{}&\omega_1(1+2\Omega)\frac{d(d-1)mc^2}{4(2+d)\pi
V_{int}a^2}\frac{1}{r_3}\exp\left(-\frac{r_3}{\lambda}\right)-\frac{\Omega mc^2\delta({\bf r}_3)}{V_{int}}\, ,\nn\\
&{}&\mu=4,\ldots ,D\, , \ea
which, for illustrative purposes, can be presented as follows:
\ba{3.4}
&{}&T_{0}^{0(eff)}\approx \frac{mc^2\delta({\bf r}_3)}{V_{int}}\left[1-\frac{d(1+2\Omega)}{2(d-2+d\omega_1)}\right]\, , \\
&{}&T_{\alpha}^{\alpha(eff)}\approx-\frac{mc^2\delta({\bf r}_3)}{V_{int}}\frac{d(1+2\Omega)}{2(d-2+d\omega_1)}\, ,\label{3.5}\\
&{}&\alpha=1,2,3\, ,\nn\\
&{}&T_{\mu}^{\mu(eff)}\approx\frac{mc^2\delta({\bf r}_3)}{V_{int}}\, \frac{d\omega_1-2(d-2)\Omega}{2(d-2+d\omega_1)}\label{3.6}\, ,\\
&{}&\mu=4,\ldots ,D\, ,\nn
\ea
where we have replaced the rapidly decreasing exponential function by the delta function:
\be{3.7} \frac{1}{r_3}\exp\left(-\frac{r_3}{\lambda}\right) \; \rightarrow\; 4\pi\lambda^2 \delta({\bf r}_3)\, . \ee
These equations shows that the effective energy density and pressures of the gravitating object depend on the parameter $\Omega$, which defines the equation of state for
the gravitating mass in the internal space. Moreover, this mass acquires the effective relativistic pressure $\hat p_{0}^{(eff)}=-T_{\alpha}^{\alpha (eff)}$ in the
external/our space. This is the crucial point. First, it is hardly possible that ordinary astrophysical objects, such as our Sun, have relativistic pressure. Second,
formulae obtained above are suitable for any gravitating mass, and not only for astrophysical objects. For example, we can apply these expressions to a system of
non-relativistic particles forming, e.g., a perfect fluid. Then, each of these particles is covered by the coat and this coat accompanies the moving particle. It is well
known that for such perfect fluid the momentum crossing the elementary spatial area $dx^{\gamma}\wedge dx^{\delta}$ per unit time is given by $T^{\alpha\beta}
\varepsilon_{\beta|\gamma\delta|}dx^{\gamma}\wedge dx^{\delta}\, ,\; \alpha,\beta,\gamma = 1,2,3 $, where $T^{ik}$ is the energy-momentum tensor of this system.
Obviously, $T^{\alpha\beta (eff)}$ must be included in the total expression for the energy-momentum tensor. Therefore, the non-relativistic particles have relativistic
momentum crossing any spatial area. Of course, it contradicts the observations. It can be easily seen that the equality $\Omega =-1/2$ is the only possibility to achieve
$\hat p_{0}^{(eff)}=0$ for our model. It means that the bare gravitating mass should have tension with the equation of state $\hat p_1=-\hat \varepsilon /2$ in the
internal space. Then, the effective and bare energy densities coincide with each other and the gravitating mass remains pressureless in our space. In the internal space
the gravitating mass still has tension with the parameter of state $-1/2$. Therefore, to be in agreement with observations, the presence of tension is a necessary
condition for the considered model. However, we still do not know a physically reasonable explanation for the origin of tension for non-relativistic gravitating objects.


\section{Conclusion}

In this paper, we investigated the viability of Kaluza-Klein models with spherical compactification of the internal space. Here, the internal sphere has an arbitrary
number of dimensions. First, we considered the famous gravitational experiments (the deflection of light and the time delay of radar echoes) in the Solar system. To do
it, we introduced a gravitating source with the dust-like equation of state $\hat p_0=0$ in the external space and an arbitrary equation of state $\hat
p_1\approx\Omega\rho c^2$ in the internal space. The perturbed (up to $O\left(1/c^2\right)$) metric coefficients were found from the Einstein equations. For the external
space, these coefficients consist of two parts: the standard general relativity expressions plus the admixture of the Yukawa interaction. The Yukawa interaction arises
only in the case when the background matter satisfies some condition (see the inequality \rf{2.26}) for the parameter of the equation of state in the internal space
which is equivalent to the condition of the internal space stabilization. From the cosmological point of view, such stabilization was considered in papers
\cite{exci,Zhuk} (see also the appendix in \cite{EZ4}). The stabilization takes place if conformal excitations of the internal spaces (referred to as gravexcitons
\cite{exci} or radions) acquire the positive mass squared. In our paper, we have shown that the mass of the Yukawa interaction is exactly defined by the mass of the
gravexciton/radion. In the Solar system, the Yukawa mass is big enough for dropping the admixture of this interaction and getting very good agreement with the
gravitational tests for any value of $\Omega$.

However, our subsequent investigation showed that the gravitating body acquires the effective relativistic pressure in the external/our space. It means that any system
of non-relativistic particles may have the relativistic momentum crossing any spatial area. Of course, it contradicts the observations. Finally, we have demonstrated
that the value $\Omega=-1/2$ (i.e. tension!) is the only possibility to avoid this problem.

\section*{ACKNOWLEDGEMENTS}
This work was supported in part by the "Cosmomicrophysics-2" programme of the Physics and Astronomy Division of the National Academy of Sciences of Ukraine. A.~Zh.
acknowledges the hospitality of the Abdus Salam International Centre for Theoretical Physics, where a part of this  work was carried out.




\begin{thebibliography}{99}

\bibitem{Landau}
L.D. Landau and E.M. Lifshitz, {\it The Classical Theory of Fields, Fourth Edition: Volume 2 (Course of Theoretical Physics Series)}, (Pergamon Press, Oxford, 2000).
\bibitem{EZ4}
M. Eingorn and A. Zhuk, Phys. Rev. D {\bf 83}, 044005 (2011); (arXiv:gr-qc/1010.5740).
\bibitem{EZ5}
M. Eingorn, O. de Medeiros, L. Crispino  and A.~Zhuk, Phys. Rev. D {\bf 84}, 024031 (2011); (arXiv:gr-qc/1101.3910).
\bibitem{EZnonlin}
M. Eingorn and A. Zhuk, Phys. Rev. D {\bf 84}, 024023 (2011); (arXiv:gr-qc/1104.1456).
\bibitem{EZf(R)solitons}
M. Eingorn and A. Zhuk,  Phys. Rev. D {\bf 85}, 064030 (2012);  (arXiv:gr-qc/1112.1539).
\bibitem{EZ3}
M. Eingorn and A. Zhuk, Class. Quant. Grav. {\bf 27}, 205014 (2010); (arXiv:gr-qc/1003.5690).
\bibitem{EZ6}
M. Eingorn and A. Zhuk,  Phys. Lett. B {\bf 713}, 154-159 (2012);  (arXiv:gr-qc/1201.1756).
\bibitem{TF}
J. Traschen and D. Fox, Tension Perturbations of Black Brane Spacetimes, Class. Quant. Grav. {\bf 21} (2004) 289-306; (arXiv:gr-qc/0103106).
\bibitem{TZ}
P.K. Townsend and M. Zamaklar, The first law of black brane mechanics, Class. Quant. Grav. {\bf 18} (2001) 5269-5286; (arXiv:hep-th/0107228).
\bibitem{HO}
T. Harmark and  N.A. Obers, General Definition of Gravitational Tension, JHEP {\bf 0405} (2004) 043-059; (arXiv:hep-th/0403103).
\bibitem{TK}
D. Kastor and J. Traschen, Stresses and Strains in the First Law for Kaluza-Klein Black Holes, JHEP {\bf 0609} (2006) 022-039; (arXiv:hep-th/0607051).
\bibitem{ChEZ1}
A. Chopovsky, M. Eingorn and A. Zhuk, {\it Weak field limit of Kaluza-Klein models with spherical compactification: problematic aspects\/}; (arXiv:gr-qc/1107.3387).
\bibitem{ChEZ2}
A. Chopovsky, M. Eingorn and A. Zhuk, Phys. Rev. D {\bf 85}, 064028 (2012); (arXiv:gr-qc/1107.3388).
\bibitem{exci}
U. G\"unther and A. Zhuk, Phys. Rev. D {\bf 56}, 6391 (1997); (arXiv:gr-qc/9706050).
\bibitem{FR}
P.G.O. Freund and M.A. Rubin, Phys. Lett. {\bf 97B}, 233 (1980).
\bibitem{Zhuk}
A. Zhuk, {\it Conventional cosmology from multidimensional models}, in {\it Proceedings of the 14th International Seminar on High Energy Physics ''QUARKS-2006''}, (INR
Press, Moscow, 2007; Vol.2, page 264); (arXiv:hep-th/0609126).
\bibitem{EZ2}
M. Eingorn and A. Zhuk, Class. Quant. Grav. {\bf 27}, 055002 (2010); (arXiv:gr-qc/0910.3507).








\end{thebibliography}
\end{document}